\documentclass[aps,prb,twocolumn,amsmath,showpacs]{revtex4-1} 

\usepackage{color,graphicx}
\usepackage{bm}

\usepackage{amsmath}
\usepackage{amsfonts}
\usepackage{amssymb}

\usepackage{subfigure}
\usepackage[latin1]{inputenc}

\usepackage{pifont}
\newcommand{\tickNo}{\hspace{1pt}\ding{55}}

\newcommand{\sand}[2]{\left\langle #1 | #2\right\rangle}

\newcommand{\ie}{{\it i.e. }}

\renewcommand{\-}{\,-\,}

\newcommand{\br}{\mathbf{r}}

\let\oldmarginpar\marginpar
\renewcommand\marginpar[1]{\-\oldmarginpar[\raggedleft\tiny #1]%
{\raggedright\tiny #1}}

\begin{document}

\title{\textbf{Particle Entanglement Spectra for Quantum Hall states on Lattices}} 

\author{Antoine Sterdyniak}
\affiliation{Laboratoire Pierre Aigrain, Ecole Normale Supérieure, 45 rue Lhomond, 75005 Paris, France}

\author{Nicolas Regnault}
\affiliation{Department of Physics, Princeton University, Princeton, NJ 08544}
\affiliation{Laboratoire Pierre Aigrain, Ecole Normale Supérieure, 45 rue Lhomond, 75005 Paris, France}

\author{Gunnar M\"oller}
\affiliation{TCM Group, Cavendish Laboratory, J.~J.~Thomson Avenue, Cambridge CB3 0HE, UK}

\date{\today}

\begin{abstract}
We use particle entanglement spectra to characterize bosonic quantum Hall states on lattices, motivated by recent studies of bosonic atoms on optical lattices. Unlike for the related problem of fractional Chern insulators, very good trial wavefunctions are known for fractional quantum Hall states on lattices. We focus on the entanglement spectra for the Laughlin state at $\nu=1/2$ for the non-Abelian Moore-Read state at $\nu=1$. We undertake a comparative study of these trial states to the corresponding groundstates of repulsive two-body or three-body contact interactions on the lattice. The magnitude of the entanglement gap is studied as a function of the interaction strength on the lattice, giving insights into the nature of Landau-level mixing. In addition, we compare the performance of the entanglement gap and overlaps with trial wavefunctions as possible indicators for the topological order in the system. We discuss how the entanglement spectra allow to detect competing phases such as a Bose-Einstein condensate.
\end{abstract}

\maketitle

\section{Introduction}

Optical lattices present unique opportunities to simulate the physics of charged particles in strong magnetic fields.\cite{Dalibard:2011p2863} While early proposals for artificial gauge fields relied on rotation to mimic the action of Lorentz forces by the Coriolis forces acting in the rotating frame,\cite{Cooper:1999p2058,Cooper:2008p250} optical lattices provide a robust experimental setting where the effect of fields can be simulated by imprinting complex phases on tunnelling elements between neighbouring sites.\cite{Jaksch:2003p91,Mueller:2004p68,Srensen:2005p58}  The most elegant schemes for generating synthetic gauge fields in atomic gases rely on the use of Berry-phases resulting from a set of internal states subject to a spatially varying optical dressing.\cite{Lin:2009p33,Cooper:2011p1498,Cooper:2011p3050} In particular, these schemes enable the generation of high densities of flux per plaquette on the underlying lattice, giving access to a regime of strong correlation where exotic topologically ordered phases can appear, including 
fractional quantum Hall liquids such as the Laughlin state\cite{Srensen:2005p58,Hafezi:2007p67} and more exotic Hall liquids which rely on the presence of the lattice.\cite{Palmer:2006p63,Moller:2009p184,Hormozi:2012p3156}

A related class of lattice-models relies on spin-orbit coupling to generate complex hopping terms in single particle tight-binding Hamiltonians. If the resulting single-particle bands are flat and have a non-zero Chern number, these systems can support states resembling fractional quantum Hall (FQH) liquids that are known as fractional Chern insulators (FCI).\cite{Neupert:2011p1803,Sun:2011p1781,Tang:2011p1780,sheng-natcommun.2.389} Fractional Chern insulators have been most convincingly shown to exhibit the same type of topological order as FQH states by analysing their particle entanglement spectra.\cite{Regnault:2011p2571}

The entanglement spectra (ES) have been initially introduced by Li and Haldane\cite{Li:2008p2861} in the context of the FQHE, stimulating an extensive range of studies.\cite{Regnault-PhysRevLett.103.016801,Zozulya-PhysRevB.79.045409,Lauchli-PhysRevLett.104.156404,Thomale-PhysRevLett.104.180502,Regnault:2011p2571,Thomale-PhysRevB.84.045127,Hermanns-PhysRevB.84.121309,Chandran-PhysRevB.84.205136,Ardonne-PhysRevB.84.205134,Sterdyniak-1367-2630-13-10-105001,Liu-PhysRevB.85.045119,Sterdyniak:2012p2886,Dubail:2012p2980,Rodriguez-PhysRevLett.108.256806,Schliemann-PhysRevB.83.115322,Alba-PhysRevLett.108.227201} They have also been studied and applied to several other systems including spin systems,\cite{Calabrese-PhysRevA.78.032329,Pollmann-1367-2630-12-2-025006,Pollmann-PhysRevB.81.064439,Thomale-PhysRevLett.105.116805,Lauchli-PhysRevB.85.054403,Yao-PhysRevLett.105.080501,Cirac-PhysRevB.83.245134,Peschel-0295-5075-96-5-50006,Huang-PhysRevB.84.125110,Lou-PhysRevB.84.245128,Schliemann-2012arXiv1205.0109S} as well as topological insulators,\cite{Fidkowski-PhysRevLett.104.130502,Prodan-PhysRevLett.105.115501,Turner-PhysRevB.82.241102} Bose-Hubbard models\cite{Liu-PhysRevA.83.013620} or complex paired superfluids.\cite{Dubail-PhysRevLett.107.157001} The ES corresponds to the spectrum of the reduced density matrix of the system groundstate when one cuts the system into two parts. The system partition can be performed in different manners such as a real space, momentum or particle space partition. Each cut can unveil different aspects of the state that is probed. In the case of the FQHE, the ES are related to the bulk or edge excitations. As these features characterize the given phase, the ES acts as a fingerprint of the system that only requires knowledge of the groundstate wavefunction.

For the above reasons, the ES were found to be particularly well suited as a tool to characterize FCI states as they only require knowledge of the ground state wave function which can be obtained numerically for small model systems.\cite{Regnault:2011p2571} Hence, it was possible to establish a detailed correspondence between the entanglement spectra of FCI with those of fractional quantum Hall states.\cite{Bernevig-2012PhysRevB.85.075128} In particular, it has recently been shown\cite{Bernevig-2012arXiv1204.5682B} that the ES is able to distinguish between a Laughlin-like state and a charge-density wave state (CDW). However, an important difference is that for fractional quantum Hall states, very accurate analytic many-body trial wavefunctions capturing the essence of these strongly correlated quantum liquids are known,\cite{Laughlin:1983p301,Jain:1989p294,Moore:1991p165,Read:1999p194,Bonderson-PhysRevB.78.125323} so a very detailed understanding of the topological order and the fundamental excitations in FQH systems has been achieved. We note that mappings of FQH wavefunctions onto topological flat bands\cite{Qi:2011p1804,Wu:2012p3204,Scaffidi:2012p3333,2012arXiv1207.4439W,2012arXiv1207.5587L} have recently led to some encouraging results, including considerable overlaps with FCI eigenstates\cite{Wu:2012p3204,Scaffidi:2012p3333} and the demonstration of an analytic continuation between these systems.\cite{Scaffidi:2012p3333,2012arXiv1207.5587L}

In this paper, we focus on quantum Hall states on lattices with a homogeneous density of gauge flux, as this gives us access to a lattice based system where quantum Hall states can be understood both in terms of entanglement spectra and many-body trial wavefunctions. As opposed to FCIs, quantum Hall states on lattices admit a continuum limit which is known to be exactly the usual FQHE. We deploy exact numerical diagonalization on a square lattice with periodic boundary conditions as the main tool for our investigation. Our study provides an independent identification of the incompressible phases on the lattice as fractional quantum Hall states by analysing the counting of quasiparticle states in the particle entanglement spectrum. To obtain the expected state counting, we establish the correspondence between the momentum sectors of a lattice-based system with periodic boundary conditions and the corresponding continuum problem on the torus. We find that for small enough particle density $n$ 
per lattice site, the counting of the continuum problem is accurately reproduced. In particular,
 this allows us to mount evidence in favour of the non-Abelian Moore-Read state as the groundstate of bosons with repulsive three-body interactions at filling factor $\nu=1$ on the lattice. Finally, we study how the entanglement gap, defined as the distance between the eigenvalues of the entanglement spectrum related to the universal property of the bulk excitations and the non-universal states at higher entanglement energy, relates to the magnitude of the overlap with the respective trial wavefunctions describing the target phase. This establishes the entanglement gap as a good proxy for the overlap, and hence for the stability of topological order.

The structure of this paper is as follows: In section \ref{sec:model}, we introduce the model Hamiltonian for bosons on lattices in the presence of magnetic fields and with periodic boundary conditions. Section \ref{sec:Torus} is devoted to a discussion of the corresponding continuum FQHE problem on the torus, where we introduce the many-body wavefunctions of the target phases which we explore in this paper, namely the Laughlin state (section \ref{sec:LaughlinOnTorus}) and the Moore-Read state (section \ref{sec:MROnTorus}). We then review the definition and characteristics of entanglement spectra in section \ref{sec:EntanglementSpectra}, and present a detailed analysis of the particle entanglement spectra of our two target phases in section \ref{sec:TargetPhases}. In the case of the Laughlin state, we describe how certain features of the ES can be used as a probe to detect a competing Bose-Einstein condensed phase. Finally, our results are summarized in section \ref{sec:Conclusions}.

\section{Model}
\label{sec:model}

We study the physics of interacting bosons on a two-dimensional square lattice in a homogeneous magnetic field applied in the direction perpendicular to the lattice. This problem is described by the Bose-Hubbard model with minimal coupling to a gauge field by Peierls' substitution. We further assume the presence of on-site two-body interactions of strength $U$ and three-body interactions of strength $V$, yielding the many-body Hamiltonian
\begin{align}
\label{eq:Hamiltonian}
\mathcal{H} = & - t \sum_{\langle \br,\br' \rangle} \left( e ^{i A_{\br\br'}}\hat a_{\br}^{\dagger}\hat a_{\br'} + h.c. \right)  \\
 & + U \sum_{\br} \hat  a_{\br}^\dagger \hat a_{\br}^\dagger\hat a_{\br} \hat  a_{\br} +  V  \sum_{\br} \hat  a_{\br}^\dagger \hat a_{\br}^\dagger a_{\br}^\dagger\hat a_{\br}\hat a_{\br} \hat  a_{\br}. \nonumber
\end{align}
Here, $\langle \br,\br' \rangle$ denotes neighbouring lattice sites $\br=(x,y)$, $\hat a_{\br}^{(\dagger)}$ are bosonic annihilation (creation) operators, and $A_{\br\br'} =\int_\br^{\br'}\mathbf{A}\cdot\mathrm{d}\mathbf{l}$ are Aharonov-Bohm phases deriving from the coupling to the underlying vector potential $\mathbf A$. We adopt units such that the lattice spacing is one, and positions $(x,y)$ can be indicated as integers.

Experimentally, bosonic Hubbard models can be engineered in optical lattices systems,\cite{Jaksch:1998p2391} where gauge potentials can be simulated by a range of different setups.\cite{Jaksch:2003p91,Mueller:2004p68,Srensen:2005p58,Lin:2009p33,Cooper:2011p1498,Cooper:2011p3050} Two-body interactions can be conveniently introduced by Feshbach resonances,\cite{Inouye:1998p3051,Greiner:2002p53} and there are proposals for three-body interactions based on strong three-particle losses.\cite{Syassen:2008p3058,Daley:2009p3057} Given this pace of progress in simulating Hubbard models, Hamiltonians of the form (\ref{eq:Hamiltonian}) may be realizable within the near future.

For our numerical exact diagonalization calculations, we express the Hamiltonian (\ref{eq:Hamiltonian}) for an ensemble of $N$ bosons on a square lattice with $N_s = L_x L_y$ sites in the presence of $N_{\phi}$ flux quanta and with periodic boundary conditions in both the $x$- and $y$-direction. This set-up corresponds to a field of flux density $n_{\phi} = N_{\phi}/N_s$, which we choose to describe in the Landau gauge 
\begin{equation}
\mathbf A = 2\pi n_\phi x \mathbf e_y,
\end{equation}
such that momentum in the $y$-direction is a conserved quantity. Due to the concurrence of a periodic lattice potential, periodic boundary conditions, and the presence of a magnetic field, this translational symmetry is reduced\cite{Kol:1993p82} to possible momenta of $k_y=2m\pi/K_y^\text{max}$, $m=0,\ldots,K_y^\text{max}-1$ with the maximal momentum index given by the greatest common denominator
\begin{equation}
\label{eq:folding}
K_y^\text{max}  = \text{gcd}(N_\phi,L_y).
\end{equation}
This can be simply explained by applying Blochs theorem to magnetic unit cells enclosing an integer number of flux quanta. Due to the reduced symmetry, orbitals are labelled by a sublattice index $s$ for the $y$-position inside the magnetic unit cell in addition to the momentum $k_y$.
Both interaction terms in (\ref{eq:Hamiltonian}) conserve this Landau-momentum, so the Hamiltonian is block-diagonal and we construct the eigenstates in the Fock space given by $| \alpha \rangle = \prod_\zeta (\hat a_\zeta^{\dagger})^{n_\zeta(\alpha)} | 0 \rangle$, where $\zeta=(x,k_y,s)$ denotes the set of single-particle quantum numbers.

We should stress that this set of states does not imply a projection to the lowest Landau-level. Instead, it includes all of the bands of the fractal single-particle spectrum known as the Hofstadter butterfly,\cite{Hofstadter:1976p69} such that Landau-level mixing is part of the model. The equivalent Landau level filling, i.e.~the particle density with respect to the number of states in the lowest band, is given by $\nu = N/N_{\phi}$.

\section{FQH on the torus}
\label{sec:Torus}

We now briefly describe some properties of the FQHE in the torus geometry. We consider a torus spanned by $\mathbf{L}_1 = L'_x \mathbf{e}_x$ and $\mathbf{L}_2 = L'_y \mathbf{e}_y$, where $\mathbf{e}_x$ and $\mathbf{e}_y$ are two perpendicular unit vectors. Given a setting where the torus is pierced by $N_{\phi}$ flux quanta, we have $L'_xL'_y = 2\pi l_B^2 N_{\phi}$, where $l_B$ is the magnetic length. The Hamiltonian is given by
\begin{eqnarray}
\label{eq:HamOnTorus}
H = \frac{1}{2m} \sum_i^N\mathbf{\Pi}^2_i + U\sum_{i < j}\tilde{\delta}( \mathbf{r}_i-\mathbf{r}_j ) \nonumber \\ + V\sum_{i < j < k }\tilde{\delta}(\mathbf{r}_i-\mathbf{r}_j)\tilde{\delta}(\mathbf{r}_j-\mathbf{r}_k)
\end{eqnarray}
where $\mathbf{\Pi}_i = -i\hbar\nabla_i - e \mathbf{A}(\mathbf{r}_i)$ is the canonical momentum of particle $j$ in the presence of a magnetic field. Since we use periodic boundary conditions, the delta function is defined as $\tilde{\delta}(\mathbf{r}) = \sum_{n,m} \delta(\mathbf{r}+ n\mathbf{L}_1 +m\mathbf{L}_2)$.

In the Landau gauge ($\mathbf A = 2\pi n_\phi x \mathbf e_y$), the one-particle orbital in the lowest Landau level with momentum index $j = 0 \dots N_{\phi} - 1$ is given by
\begin{eqnarray}
\label{eq:TorusOribtals}
\phi_{j}(x,y) = \frac{1}{(\sqrt{\pi}L'_yl_B)^{1/2}} \exp\left[-\frac{x^2}{2l_B^2}\right]\nonumber \\ \vartheta\begin{bmatrix} \frac{j}{N_{\phi}}\\0\\                                                                                                                                                                                                                                                                                                                                                                                                                                                                               \end{bmatrix}
\left(\frac{N_{\phi}}{L'_y}(y-ix)\right.\left| i\frac{L'_x}{L'_y}\right),
\end{eqnarray}

where $\vartheta\begin{bmatrix} a\\b\\ \end{bmatrix}(z|\tau) =\sum_n e^{i\pi\tau (n+a)^2+2i\pi (n+a)(z+b)}$ are the generalized Jacobi theta functions. This model is the continuous version of (\ref{eq:Hamiltonian}).

As for the lattice, the momentum in the $y$-direction is a conserved quantity and the $N$-body Hamiltonian is block diagonal with respect to the total momentum $\mathcal{K}_y^T = \sum_i j_i \textrm{mod} N_{\phi}$. Note that in the case of the lattice, the total momentum $\mathcal{K}_y$ is defined modulo $K_y^\text{max}$.

\subsection{Laughlin state}
\label{sec:LaughlinOnTorus}

In the lowest Landau level, when only two-body interactions are present \ie $V=0$ in equation (\ref{eq:HamOnTorus}), the Laughlin state is the densest zero energy groundstate.\cite{Laughlin:1983p301} A hallmark of this phase is its twofold groundstate degeneracy. For a finite size system with $N_\phi=2N$ the two groundstates can be found at momenta of $\mathcal{K}^T_y=0,\,N$. 
They are given by:
\begin{eqnarray}
\label{eq:Laughlin}
\Psi(z_1,\ldots,z_N) = f_\text{rel}(z_1,\ldots,z_N)F_{c.m.}(Z)e^{-\frac{1}{2}\sum_ix_i^2 /l_B^2}, \nonumber \\
\end{eqnarray}
where $F_{c.m.}$ is a center of mass wavefunctions that depends only on the center of mass coordinate $Z = \sum_iz_i$, while $f_\text{rel}$ is the wavefunction describing the relative motion. On a rectangular torus of size ($L'_x\times L'_y$), we have
\begin{equation}
\label{eq:Laughlin-rel}
 f_\text{rel}=\prod_{i<j} \vartheta \bigg[\begin{array}{c} \frac{1}{2}\\ \frac{1}{2} \\                                                                                                                                                                                                                                                                                                                                                                                                                                                                               \end{array}\bigg]
\left(\frac{z_i-z_j}{L'_y}\bigg|i\frac{L'_x}{L'_y}\right)^2.
\end{equation}

Due to the symmetry under translations of the center of mass, the center of mass wavefunction at $\nu =1/2$ is two-fold degenerate\cite{Haldane:1985p2786} and is given by
\begin{equation}
\label{eq:CM}
 F_{c.m.}(Z) = \vartheta \bigg[\begin{array}{c} \frac{l}{2}+\frac{N_{\phi}-2}{4}\\ \frac{2-N_{\phi}}{2} \\                                                                                                                                                                                                                                                                                                                                                                                                                                                                               \end{array}\bigg]
\left(\frac{2Z}{L'_y}\bigg|2i\frac{L'_x}{L'_y}\right),
\end{equation}
where $l=0,1$ indexes the two degenerate wavefunctions.

\subsection{Moore-Read state}
\label{sec:MROnTorus}

In the lowest Landau level, the Moore-Read state\cite{Moore:1991p165} is the densest zero-energy ground state of the hardcore three-body interactions, given by Eq.~(\ref{eq:HamOnTorus}) with $U = 0$. It embodies the physics of a chiral $p$-wave superconductor of composite fermions,\cite{ReadGreen00,MollerSimon08} which can be cast in the real-space form of a BCS paired state in terms of the pair-wavefunction $1/(z_i-z_j)$. An equivalent expression can be found for the torus, where the ground state is three fold-degenerate. On a rectangular torus, these groundstates can be found at $k_y$-momenta $\mathcal{K}^T_y=\{0 ,0,\frac{N}{2}\}$. For our purposes, it is most useful to obtain the Moore-Read trial wavefunctions from the Laughlin state by using the Cappelli formula,\cite{Cappelli:2001p3153} which relates the Moore-Read state of $N$ particles to two independent layers of Laughlin $\nu=1/2$ states with half the number of particles. Using notations for a sphere or disk:
\begin{multline}
\Psi_{\rm Pf}(z_1,\ldots,z_N) =\\
\mathcal{S}\left(\prod_{i<j=2}^{N/2} (z_i-z_j)^2 (z_{\frac{N}{2}+i}-z_{\frac{N}{2}+j})^2 \right).
\end{multline} 
where $\mathcal{S}$ is the symmetrization operator. On the torus, the Laughlin state is two fold-degenerate while the Moore-Read state is three fold-degenerate. Similar to what happens for the quasihole states on the disk or sphere geometry, the symmetrization induces linear dependencies. Here, one can write four states:
\begin{eqnarray}
\Psi^{\rm{Pf}}_{K^T_y}(z_1,...,z_N)=\mathcal{S} ( &&\Psi^{Lg}_{K^T_{y_1}}(z_1,...,z_{N/2})\\
 &&\times \Psi^{Lg}_{K^T_{y_2}}(z_{N/2+1},...,z_N) )\nonumber
\end{eqnarray}
with $K^T_{y_1}, K^T_{y_2}$ being equal to one of the degenerate Laughlin states with $K^T_{y}=0$ or $K^T_{y}=N'=\frac{N}{2}$. The total momentum $K^T_y = K^T_{y_1} + K^T_{y_2} \textrm{mod}\, N_{\phi}$  perfectly matches the one of the Moore-Read state: taking $K^T_{y_1}=K^T_{y_2} =0$ yields a first $K^T_y = 0$ MR state, taking $K^T_{y_1}=K^T_{y_2} = N'$ again yields $K^T_y = 0$ and finally taking $K^T_{y_1}=0$ and $K^T_{y_2} = N'$ one obtains the remaining MR ground state at $K^T_y = N'=N/2$ state (Choosing $K^T_{y_1}=N'$ and $K^T_{y_2} = 0$ is equivalent due to the symmetrization and does not yield an additional state).

\section{Particle Entanglement Spectra}
\label{sec:EntanglementSpectra}

The entanglement spectrum gives access to many of the spectral properties of the system which are encoded in the groundstate wavefunction.\cite{Li:2008p2861} It is defined from the reduced density matrix of a subsystem resulting from the partition of the system into two (or more) parts $A$ and $B$. For the particle entanglement spectrum (PES) this partition consists in distributing the particles into two subgroups ($A$ and $B$) while keeping the geometry unchanged.\cite{PhysRevLett.106.100405}  The reduced density matrix $\rho_A = \mathrm{Tr}_B \rho$, obtained by tracing out the $N_B$ particles in the $B$ partition,  yields the entanglement spectrum by diagonalising and classifying the resulting eigenstates according to the symmetries of the problem. This process is equivalent to a Schmidt decomposition of the original many-body state into orthogonal bases for the partitions
\begin{equation}
\label{eq:defineES}
| \Psi \rangle = \sum_{\varpi}\sum_i e^{-\xi_{\varpi,i}/2} | \Psi_{\varpi,i}^A \rangle \otimes | \Psi_{\varpi,i}^B \rangle,
\end{equation}
where $\varpi$ stands for quantum numbers designating a sector of the decomposition and $i$ indexes states in each sector, and the eigenvalues $\lambda_{\varpi,i} = e^{-\xi_{\varpi,i}/2}$ of the decomposition are represented on a logarithmic scale. The vectors in Eq.~(\ref{eq:defineES}) are orthonormal i.e. $\langle \Psi_{\varpi,i}^A | \Psi_{\varpi ',j}^A \rangle=\langle \Psi_{\varpi,i}^B | \Psi_{\varpi ',j}^B \rangle=\delta_{i,j} \delta_{\varpi,\varpi '}$. The entanglement spectrum is given by plotting $\xi$'s over the relevant $\varpi$.

It has been observed that model states such as the Laughlin or MR states have a characteristic PES:\cite{PhysRevLett.106.100405} the number of non-zero eigenvalues for $\rho_A$ is identical to the number of quasihole states for a similar system with the same geometry and $N_A$ particles.  This number is usually exponentially lower than the dimension of $\rho_A$. The same features can often persist for eigenstates of realistic interaction Hamiltonians: for groundstate wavefunctions with robust topological order, one should observe a clearly defined entanglement gap, separating an ensemble of low-lying entanglement eigenvalues from non-universal eigenvalues located at higher entanglement energies. Notice that model wavefunctions can be thought of as having an infinite entanglement gap.

For the model associated to the Hamiltonian (\ref{eq:Hamiltonian}), we classify the sectors of the entanglement spectrum by 
the Landau-momentum $k_y = 0,\ldots, K_y^\text{max}-1$, i.e., $\varpi=k_y$. An equivalent classification can be chosen for the continuum fractional quantum Hall problem on the torus,\cite{Haldane:1985p2786} except that the corresponding Landau-momentum $\mathcal{K}_y^T$ can take all $N_\phi$ distinct values. To compare the entanglement spectra in these two distinct situations, we identify the momentum sectors modulo $K_y^\text{max}$. In particular, we will study the counting of the number $\mathcal{N}$ of low-lying entanglement eigenvalues below an entanglement gap. In this case, we expect the following mapping between the values for the torus $\mathcal{N}_T$ and lattice $\mathcal{N}_L$ 
\begin{equation}
\label{eq:mappingN}
\mathcal{N}^L(k_y) = \sum_{\mathcal{K}^T_y}  \delta_{k_y,(\mathcal{K}^T_y \,\mathrm{mod}\, K_y^\text{max})} \, \mathcal{N}^T(\mathcal{K}^T_y).
\end{equation}

Note that for degenerate groundstate manifolds, the PES has to be calculated for the incoherent average reduced density matrix $\rho_\text{tot}$ for the ensemble of groundstates $\{|\Psi_\alpha\rangle\}$, given by the sum
\begin{equation}
\label{eq:degenerateRho}
\rho_\text{tot} = \frac{1}{d_\text{GS}} \sum_\alpha | \Psi_\alpha \rangle \langle\Psi_\alpha |
\end{equation}
over the $d_\text{GS}$ degenerate ground states. As discussed in Ref.~\onlinecite{PhysRevLett.106.100405}, this definition yields model state PES for degenerate groundstate manifolds which recover the properties of the non-degenerate case on simply connected surfaces.

\section{Target Phases}
\label{sec:TargetPhases}

The presence of incompressible fractional quantum Hall liquids is well established for the Hamiltonian (\ref{eq:Hamiltonian}).
These states include the fractional quantum Hall liquids of the continuum problem,\cite{Srensen:2005p58,Hafezi:2007p67}  however, the presence of the lattice potential also gives rise to additional incompressible states.\cite{Palmer:2006p63,Moller:2009p184,Hormozi:2012p3156} In this paper, our aim is to establish the use of entanglement spectra for lattice based systems, so we shall focus on the states with an equivalent in the continuum case and undertake a comparison of their features.

\subsection{Laughlin State}
\label{sec:TargetLaughlin}
We begin our analysis with the Laughlin state of bosons at $\nu=1/2$, as the best investigated quantum Hall state on lattices.  We use the analytic form of the Laughlin states in the continuum (Eq.~\ref{eq:Laughlin}) and substitute the discrete lattice coordinates, such that $z=a/\ell_0 (m \mathbf{e}_x + i n \mathbf{e}_y)$ for lattice site $\mathbf{i}=(m,n)$. By virtue of the folding of momenta (\ref{eq:folding}), the two ground states (\ref{eq:CM}) now occur at $k_y=0$ and $k_y=N\,\textrm{mod}\,K_y^\text{max}$, i.e.~they may either remain at different $k$-points or are both mapped to zero momentum if $N\,\textrm{mod}\,K_y^\text{max} = 0$. In our simulations, we find that for lattice systems with sufficiently small particle density, the momenta of the numerically obtained ground states are in agreement with this prediction. In particular, there is an extended regime where a two-fold degenerate groundstate with a 
finite gap is found.\cite{Hafezi:2007p67} Hence, we can analyze the particle entanglement spectrum of the groundstate manifold according to the total density matrix (\ref{eq:degenerateRho}).

\begin{figure*}[htb]
\begin{center}
\includegraphics[width = 0.49\textwidth]{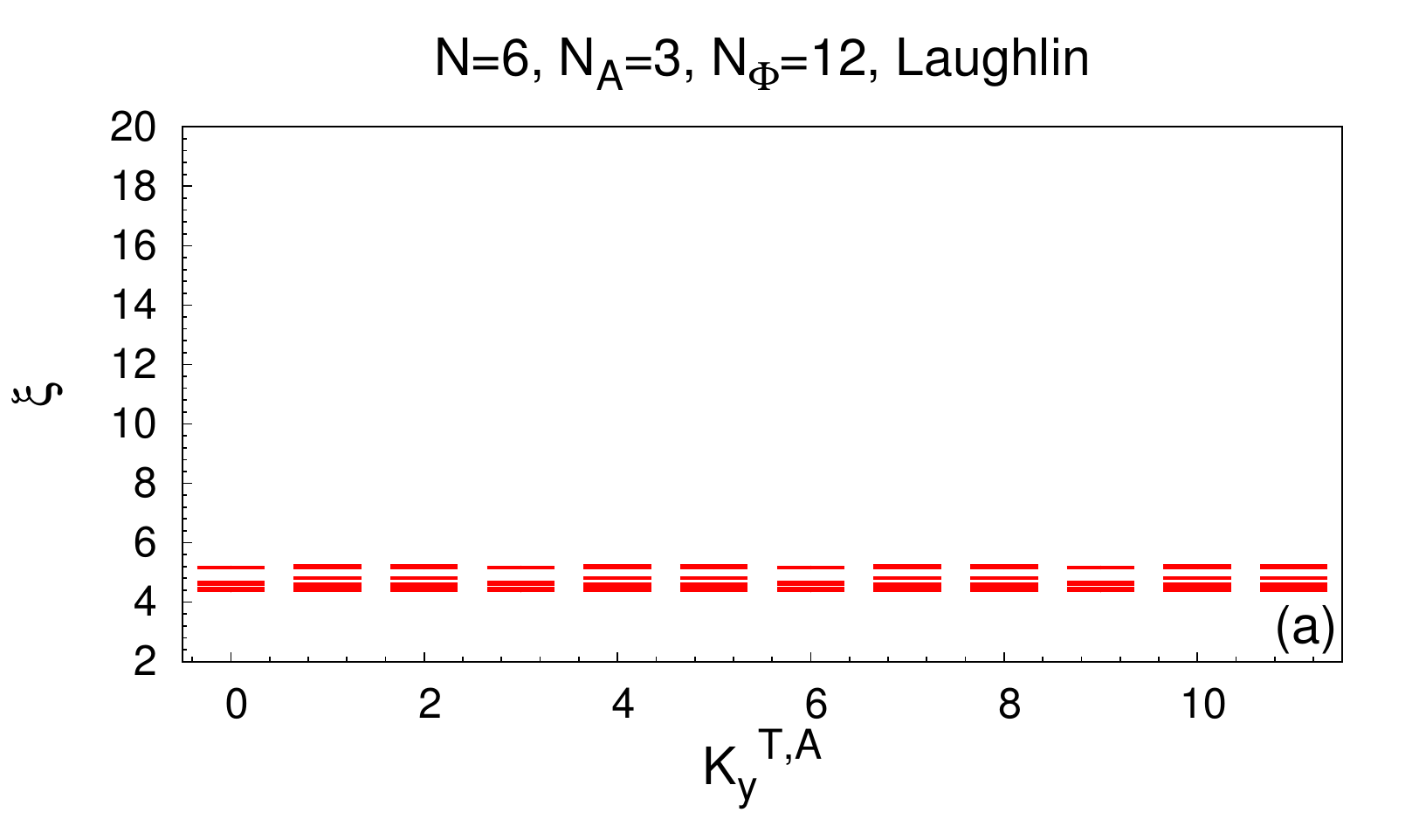}
\includegraphics[width = 0.49\textwidth]{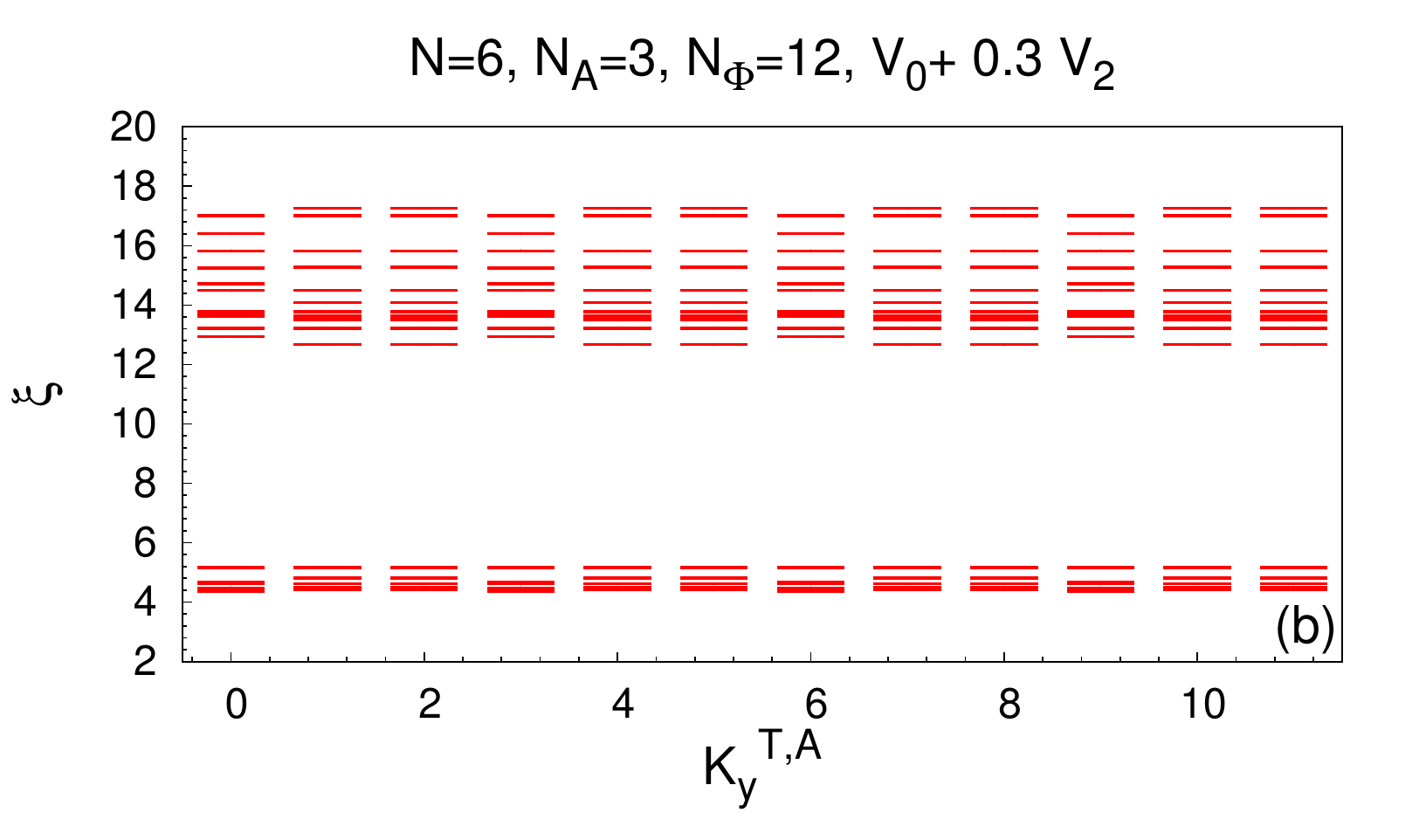}
\caption{Particle entanglement spectra for $N = 6$ bosons on the torus of unity aspect ratio at filling factor $\nu = N/N_\phi=1/2$, for a particle partition with $N_A = 3$. Left: bosons interacts through hardcore interaction. Right: bosons interacts through hardcore interaction and an additional longer range interaction.}
\label{pestoruslaughlin}
\end{center}
\end{figure*}

\begin{figure*}[htb]
\begin{center}
\includegraphics[width = 0.49\textwidth]{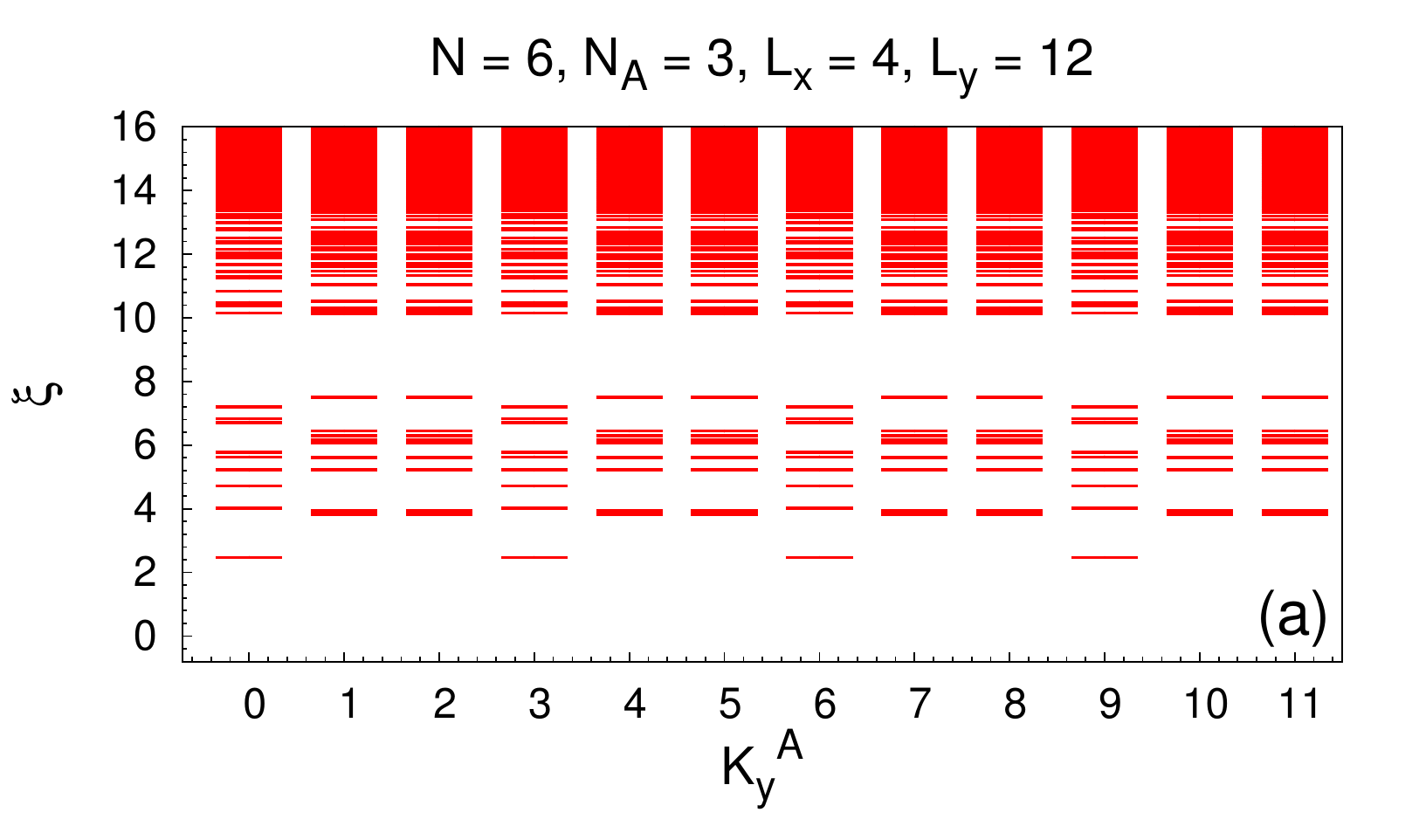}
\includegraphics[width = 0.49\textwidth]{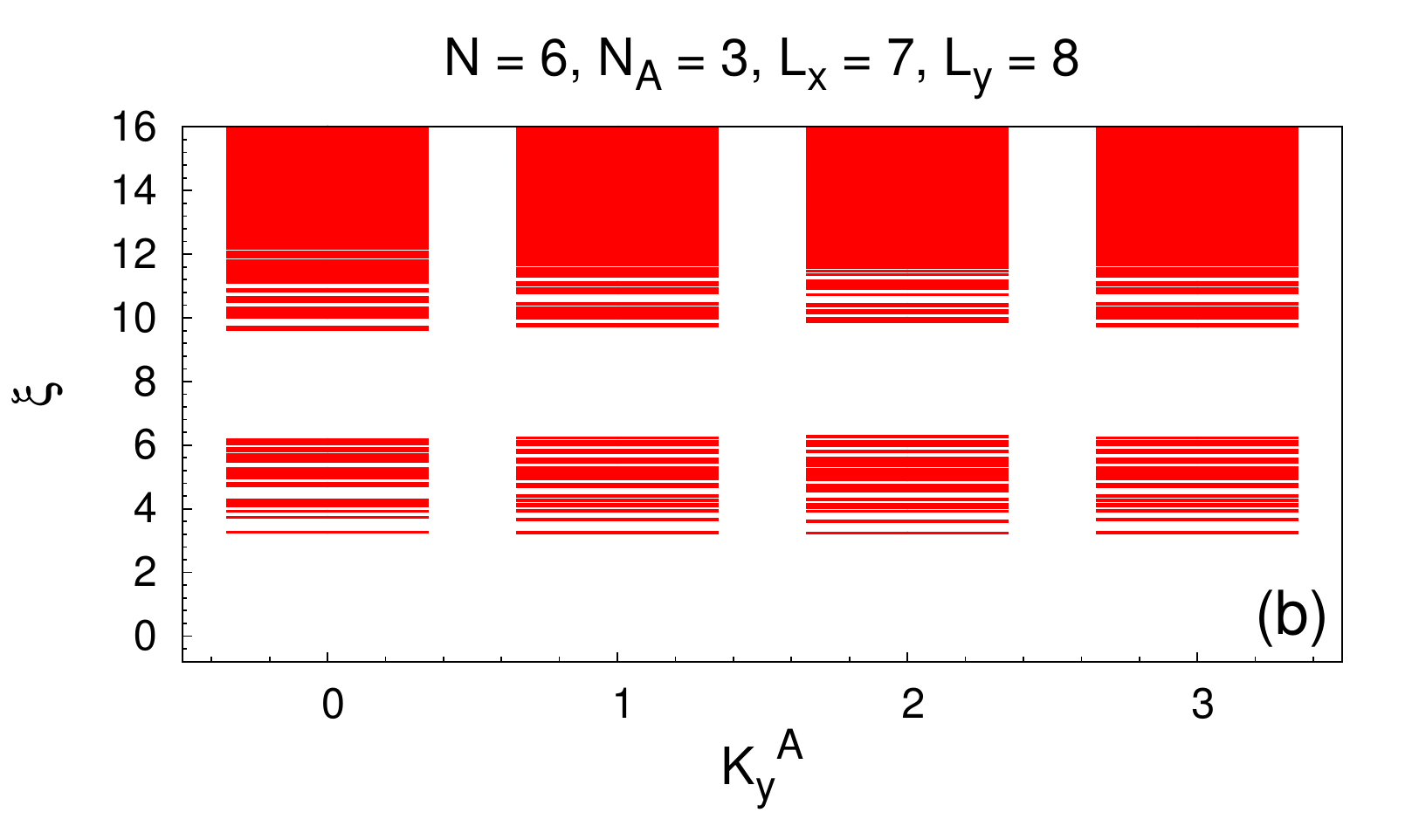}\\
\includegraphics[width = 0.49\textwidth]{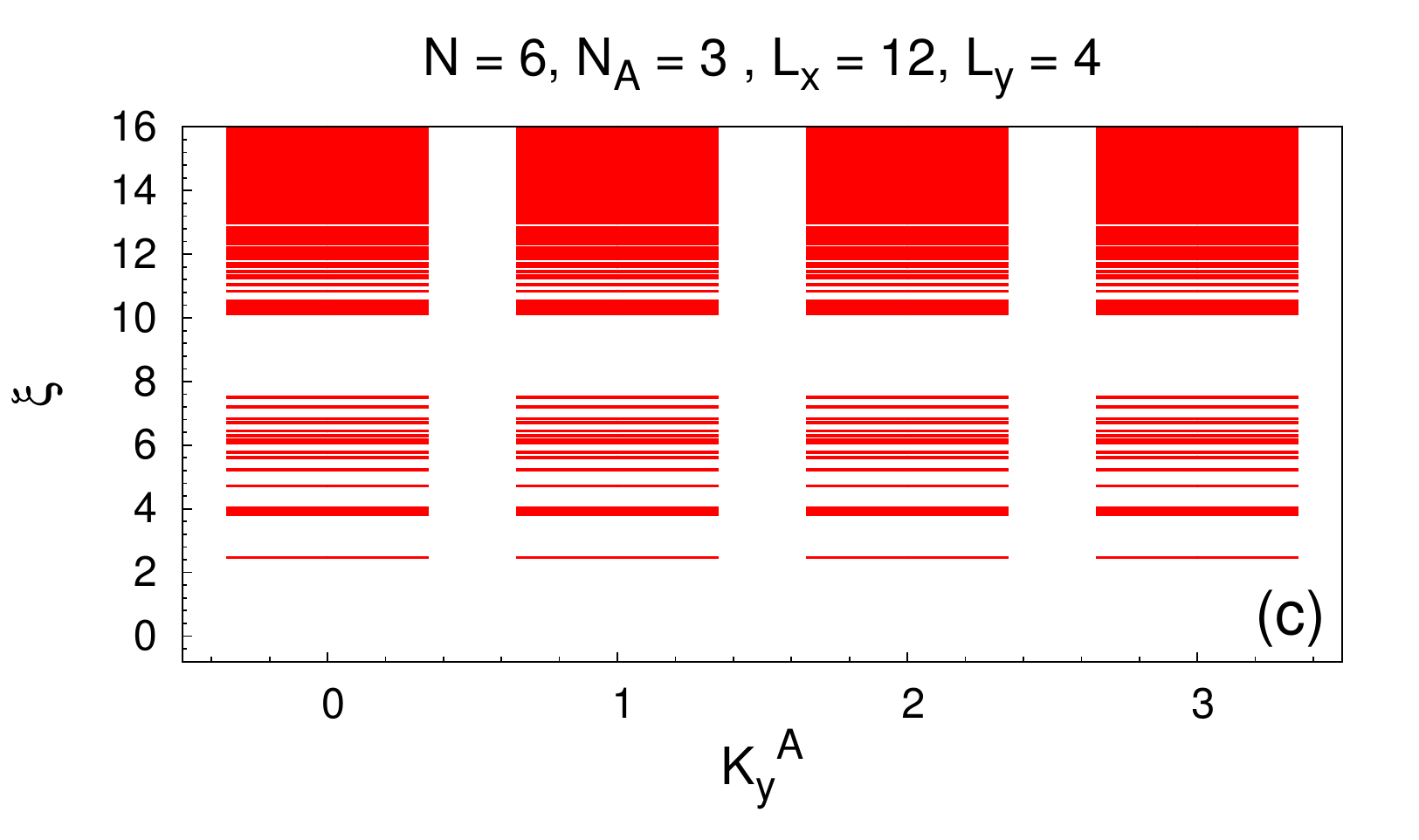}
\includegraphics[width = 0.49\textwidth]{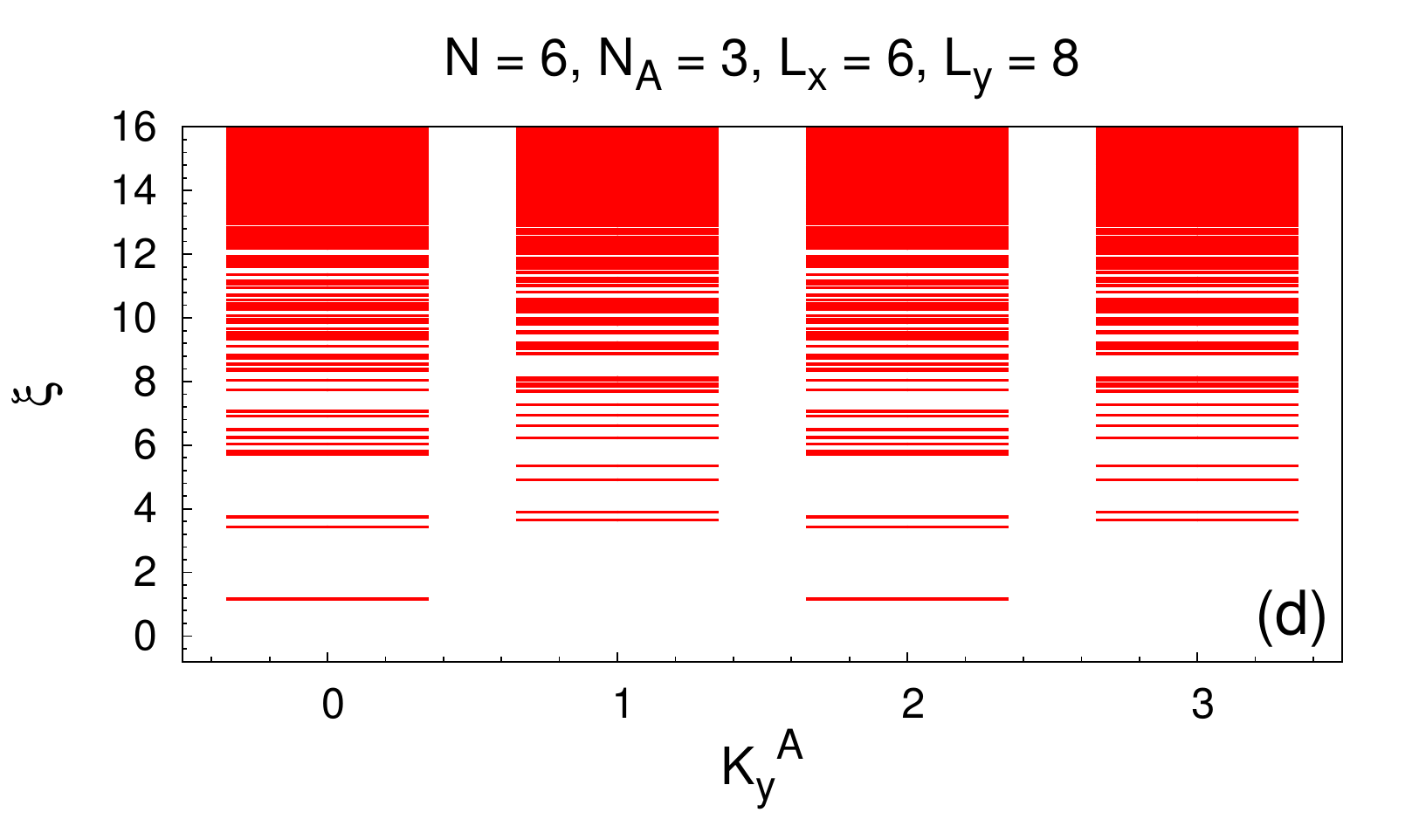}
\caption{Particle entanglement spectra for $N = 6$ bosons on a lattice at filling factor $\nu = N/N_\phi=1/2$, for a particle partition with $N_A = 3$ and for different lattice geometries. The spectra are calculated for the two-fold degenerate groundstate manifold of the Hamiltonian (\ref{eq:Hamiltonian}) with $U/t=1$ and $V=0$.}
\label{fig:ES_N6}
\end{center}
\end{figure*}

\begin{figure*}[htb]
\begin{center}
\includegraphics[width = 0.49\textwidth]{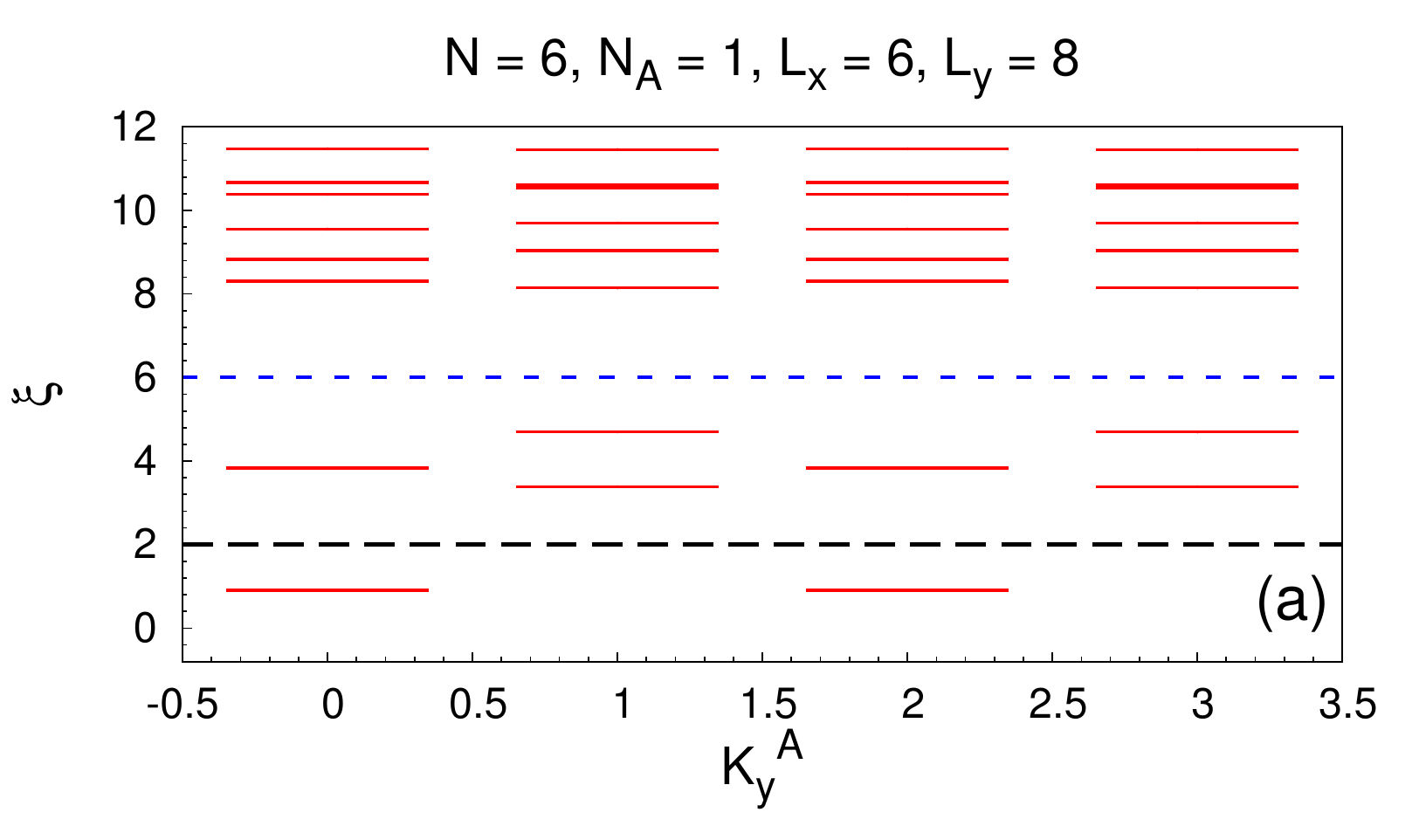}
\includegraphics[width = 0.49\textwidth]{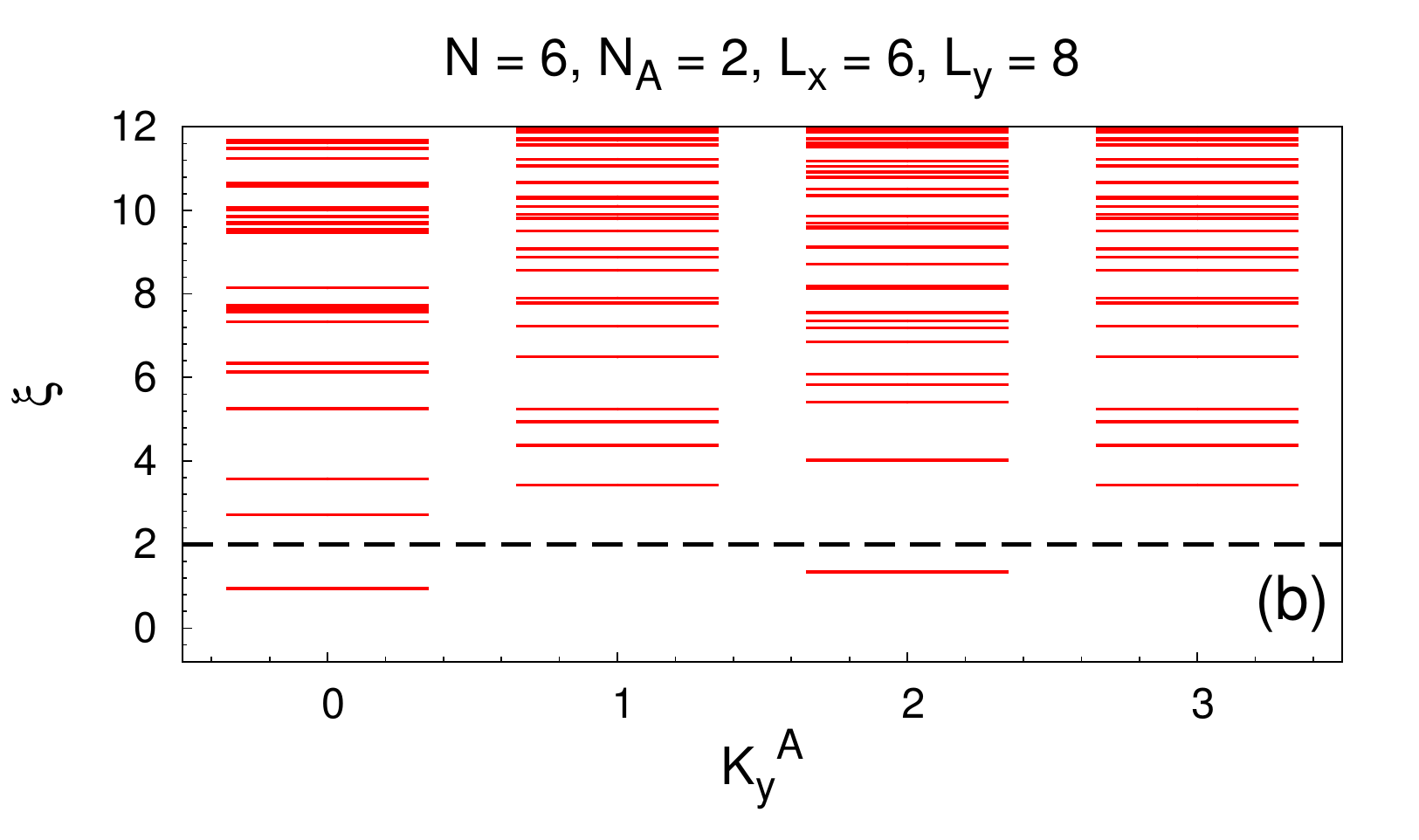}
\caption{Particle entanglement spectra for $N = 6$ bosons on a lattice of $6\times 8$ sites at filling factor $\nu = N/N_\phi=1/2$ for particle partitions with $N_A = 1$ (a) and $N_A = 2$ (b) [$N_A = 3$ shown in Fig.~\ref{fig:ES_N6}(d)]. For $N_A = 1$, the upper blue dashed line marks the entanglement gap below which the counting matches the lowest Landau level state counting of 12 states, and the lower black dashed line marks the entanglement gap that isolated the two low-lying eigenstates. For $N_A = 2$, the black dashed line marks the entanglement gap that isolated the two low-lying eigenstates.}
\label{fig:ES_N6_6x8}
\end{center}
\end{figure*}

We begin to illustrate the mapping of the entanglement spectrum on the torus to the lattice (\ref{eq:mappingN}) for a small model system with $N=6$ particles moving in the field of $N_\phi=12$ flux quanta. We first consider the PES with $N_A = 3$. Fig.~\ref{pestoruslaughlin}a shows the PES for the Laughlin state on the torus geometry The PES yields the following counting for the $12$ distinct $\mathcal{K}^T_y$-momentum sectors on the torus: $(10,9,9,10,9,9,10,9,9,10,9,9)$. Fig.~\ref{pestoruslaughlin}b displays the PES for the groundstate of the model Hamiltonian that gives rise to the Laughlin state but with an additional small contribution from a longer range interaction. In that case, the PES exhibits a entanglement gap. The counting below the gap exactly matches the one of the Laughlin state.

We now consider the lattice model. The counting of the Laughlin on the torus is reproduced exactly on a lattice of geometry $L_x = 4$ and $L_y = 12$ for $U/t=1$ and $V = 0$ [see Fig.~\ref{fig:ES_N6}(a)] as this lattice retains $K_y^\text{max} =  N_\phi = 12$. Notice that the total number of states above the gap per momentum sector is much higher in the lattice case than in the continuum model shown in Fig.~\ref{pestoruslaughlin}b. This is a consequence of the Hilbert space dimension being set by $N$ and $L_x \times L_y$ in the lattice case, and by $N$ and $N_\phi$ in the continuum case. So, our confirmation of a clear entanglement gap for such a small number of particles in the lattice model is even more remarkable.

For lattices with other aspect ratios, a folding of the momentum axis often occurs, in which case the maximum momentum is reduced. Several examples for such entanglement spectra are shown in Fig.~\ref{fig:ES_N6}(b-d). For example, in panel (b), for the geometry with $L_x = 7$, $L_y = 8$, and $K_{\max} = 4$ the use of equation (\ref{eq:mappingN}) predicts the counting $(28,28,28,28)$, which is indeed reproduced. The same 
result is also obtained for the aspect ratio of $12\times 4$ sites in panel (c). However, we do not always obtain a PES with a well defined entanglement gap. For the lattice geometry of $L_x = 6$ and $L_y = 8$, we find that no threshold value $\xi_t$ for the entanglement energy yields a clear-cut definition of $\mathcal{N}_{\xi_t}^L(k_y)$. We could speculate whether this is due to the commensurability of the number of particles with $L_x$. In such geometries, it has previously been found that CDW states can intervene.\cite{Moller:2009p184} However, there is a range of phases which may be competing with fractional quantum Hall liquids in optical flux lattices, which include bosonic condensates with symmetry breaking,\cite{Moller:2010p828} or more general supersolid phases.\cite{Moller:2010p828,Moller:2012p2646}

As a first step towards understanding the lattice which does not conform to the picture of an incompressible Laughlin state ($L_x = 6$, $L_y = 8$), we investigate several additional entanglement spectra for this system, analysing the dependency on the number of particles in the partition $A$. The results for $N_A=1$ and $N_A=2$ are shown in Fig.~\ref{fig:ES_N6_6x8}. Firstly, we note that the entanglement spectrum for $N_A=1$ carries non-trivial information for the lattice, while the corresponding continuum limit would yield a number of eigenstates which is given by the total Hilbert space dimension, i.e., the number of states in the lowest Landau level. On the lattice, we instead find that a gap opens in the entanglement spectrum above a number of low-lying states which precisely matches the number of eigenstates in the lowest Landau level. Specifically, Fig.~\ref{fig:ES_N6_6x8}(a) reveals precisely $12$ states below the gap located at about $\xi=6$. For a Laughlin state, we would expect all twelve of these states to be degenerate. Secondly, we find that there are two eigenstates which are separated from the other $10$ by a further entanglement gap located near $\xi=2$, reproducing the same number and momentum sectors $K^A_y=0$ and $K^A_y=2$ of the two low-lying states that we had observed in the PES for $N_A=3$ in Fig.~\ref{fig:ES_N6}(d). To complete our survey, we also examine the entanglement spectrum for $N_A=2$ in Fig.~\ref{fig:ES_N6_6x8}(b). Again, we find two degenerate low-lying eigenstates with an entanglement gap near $\xi=2$ and located in the same momentum sectors. This invariance of the number of entanglement eigenvalues with the number of particles in the partition is fundamentally different from the behaviour that we expect from topologically ordered phases. By contrast, the ability to absorb further particles without any change of the physical properties (i.e., the number of low-lying excitations) can be seen as an indication of the physics of Bose condensation.

To probe for the presence of a Bose condensate, we use the single particle density matrix $\rho^s_{ij}=\langle \hat a^\dagger_i \hat a_j \rangle$, calculated between lattice sites $i$, $j$. This matrix is exactly a reduced density matrix for the specific value $N_A=1$. A state with a finite condensate fraction is signalled by single a large eigenvalue $\lambda_0$ of $\rho^s$, whose magnitude scales with the system size $N$. However, we find that the groundstate wavefunction for our system with $L_x = 6$ and $L_y = 8$ and $N_\phi=12$ yields a twofold degenerate pair of largest eigenvalues $\lambda_{0}=\lambda_1\simeq 2.472$. This characteristic is known to be associated to discrete symmetry breaking in the thermodynamic limit.\cite{Moller:2010p828} We thus follow the procedure introduced in Ref.~\onlinecite{Moller:2010p828} and calculate the density matrix for symmetry-broken states which are obtained in our case by constructing superpositions formed of the two lowest lying eigenstates $|S\rangle = c_0 |\Psi_0\rangle + (1-|c_0|^2) | \Psi_1 \rangle$ that optimize the largest eigenvalue of $\rho^s_{ij}$. For the symmetry-broken state that results from superposing two states with different momenta, we find a single large density matrix eigenvalue $\lambda_0(S) = 4.4911$, corresponding to a condensed fraction of $74.8\%$ for the $N=6$ particle system (the condensate fraction rises to $95\%$ as interactions are reduced to $U=0.1t$). At the same time, the state breaks translational invariance, forming stripes running around the short cycle of the simulation cell. Similar finite size effect had previously been reported for lattices,\cite{Moller:2009p184} as well as for continuum problems of bosons.\cite{Cooper:2007p89} It is likely that properties of the particular lattice size which we discuss here are related to its flux density of precisely $n_\phi=1/4$. At this value of $n_\phi$, the single particle Hofstadter spectrum consists of a single, moderately wide band which naturally supports Bose condensation at low interaction strength. As we do not examine this question in further detail, we can only speculate whether all of the above features survive in the thermodynamic limit, in which case the phase could be considered a supersolid.\cite{Moller:2010p828} For the purpose of the current paper, we can conclude that an entanglement spectrum with few low-lying states, whose number remains invariant for different partitions of the system, is indicative of a condensed state. Cases where such eigenvalues occur in different momentum sectors are likely related to condensates with symmetry breaking. 

\begin{table*}[htb]
\begin{tabular}{cccccccccc}
\hline
$N$ & $L_x$ & $L_y$ & $k_y(\text{GS})$  & match &$\Delta$ &  $\delta$ & PES: $\{\mathcal{N_A}_L(k_y)\}$ & match & $\Delta_\xi$\\
\hline 
\hline 
4 &4 & 4& 0,0 & \tickNo&& &  &&\\
4 &6 & 4& 0,0 & \tickNo&& &  &&\\
4 & 8& 4& 0,0 & \checkmark&0.043 &5.9e-04 & 6,4,6,4 & \checkmark&3.59\\
4 & 14& 4& 0,0 & \checkmark&0.058 &1.8e-05 & 6,4,6,4 & \checkmark&7.42\\
4 & 6& 6& 0,0 & \checkmark &0.050& 0.023&  12,8 & \checkmark&4.86\\
4 & 4& 8& 0,4 & \checkmark &0.043& 5.9e-04&  3,2,3,2,3,2,3,2 & \checkmark&3.59\\
4 & 6& 8& 0,4 & \checkmark &0.066& 7.1e-04&  3,2,3,2,3,2,3,2 & \checkmark&8.1\\
4 & 7& 8& 0,4 & \checkmark &0.057 & 2.9e-10&  3,2,3,2,3,2,3,2 & \checkmark&10.9\\
4 & 8& 8& 0,4 & \checkmark &0.052 & 0& 3,2,3,2,3,2,3,2 & \checkmark&11.5\\
4 & 10& 8& 0,4 & \checkmark &0.040 & 1.4e-08&  3,2,3,2,3,2,3,2 & \checkmark&12.9\\
\hline
5 & 6 & 8 & 0,1 & \checkmark & 0.059  & 0 &  20,15 & \checkmark& 5.5\\
5 & 8 & 8 & 0,1 & \checkmark & 0.063  & 0  &  20,15&\checkmark & 9.2\\
5 & 6 & 10 & 0,5 & \checkmark & 0.067 & 0 & 4,3,4,3,4,3,4,3,4,3 & \checkmark&8.4\\
\hline
6 & 4& 12& 0,6 &\checkmark& 0.054&0 &  10,9,9,10,9,9,10,9,9,10,9,9 & \checkmark&2.63\\
6 & 12& 4& 0,2 &\checkmark& 0.054& 0 &  28,28,28,28 & \checkmark&2.63\\
6 & 6& 8& 0,2 &\checkmark & 0.0099&0.0029& $\Delta_E=0$ & \tickNo$^\dagger$ &\\
6 & 7& 8& 0,2 &\checkmark& 0.043 &8e-6 & 28,28,28,28 & \checkmark&3.5\\
\hline
\hline
\multicolumn{10}{l}{\footnotesize $^\dagger$ See Fig.~\ref{fig:ES_N6_6x8} and main text in section \ref{sec:TargetLaughlin} for a discussion.}
\end{tabular}
\caption{Properties of the particle entanglement spectra of the groundstate manifold of the Hamiltonian for $U/t = 1$ and $V = 0$ (\ref{eq:Hamiltonian}) for different model systems of $N$ bosons with $N_\phi=2N$ flux quanta. We indicate the momenta of the two degenerate groundstates $k_y(\text{GS})$ as well as the energy gap $\Delta$ and the groundstate energy splitting $\delta$. The counting of the particle entanglement spectrum is shown for the partition with $N_A = \lfloor N/2 \rfloor$ and for different lattice geometries. For both these properties, we indicate the agreement with the predictions for the Laughlin state and the entanglement gap $\Delta_\xi$.}
\label{tab:systems}
\end{table*}

Returning to our main discussion of the properties of the Laughlin state, we present a collection of the properties of the entanglement spectra in table \ref{tab:systems}, which gives an overview for several lattice geometries that we have studied.
To summarize our principal findings from these data, we have established that the counting of excitations encoded in the PES for the Laughlin states on the lattice Hamiltonian at filling factor $\nu=1/2$ agrees well with the data for the continuum problem in the lowest Landau level for small flux/particle density per plaquette. In particular, the entanglement spectra of this state show a clear entanglement gap $\Delta_\xi$. 

\begin{figure}[tb]
\includegraphics[width = 0.95 \columnwidth]{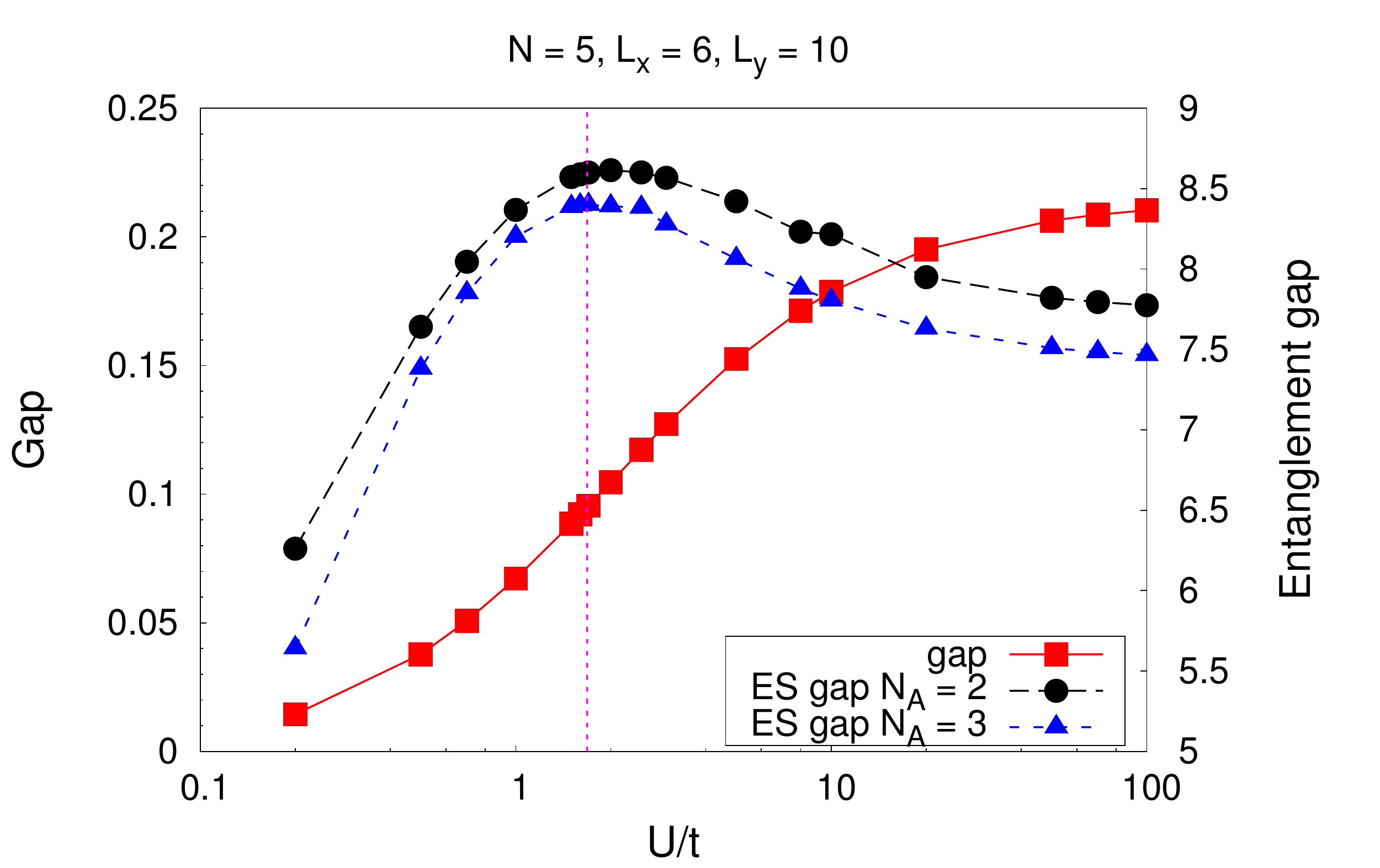}
\caption{Energy gap $\Delta$ and entanglement gap $\Delta_\xi$ in the $N_A = 2$ and in the $N_A = 3$ sectors as a function of $U$ for $\nu=1/2$, $N = 5$, $L_x = 6$ and $L_y = 10$. The vertical purple line is the band gap. Note the different offsets on the scales for $\Delta$ and  $\Delta_\xi$.}
 \label{fig:ES_gap_vs_u}
\end{figure}

Unlike the problem on the torus, where it is customary to consider the projection of the Hamiltonian into the lowest Landau level, the full lattice Hamiltonian (\ref{eq:Hamiltonian}) includes all Landau (Hofstadter) bands. Hence, we can study the effect of band mixing that occurs as a function of the interaction strength $U$. The evolution of the entanglement gap $\Delta_\xi$ with $U$ is shown in Fig.~\ref{fig:ES_gap_vs_u}, alongside the energy gap $\Delta$. Unlike the energy gap which always increases with $U$, the entanglement gap reaches a maximum value for an interaction strength of the order of the band gap and then decreases. 

\begin{figure}[htb]
\includegraphics[width = 0.95 \columnwidth]{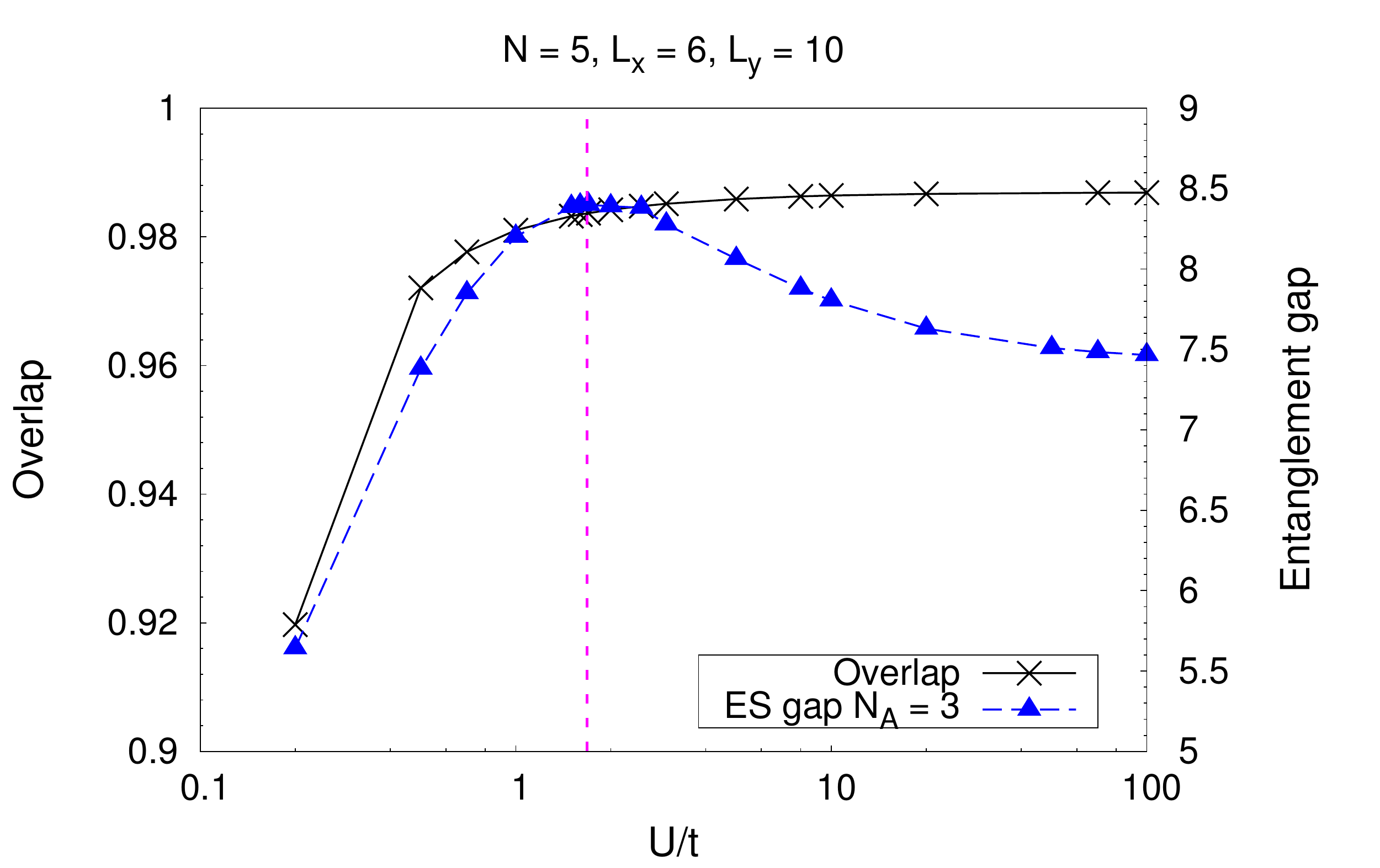}
\caption{Overlap $\mathcal{O}$ and entanglement gap $\Delta_\xi$ in the $N_A = 3$ sectors as a function of $U$ for $\nu=1/2$, $N = 5$, $L_x = 6$ and $L_y = 10$. The vertical purple line is the band gap. Note the different offsets on the scales for $\mathcal{O}$ and  $\Delta_\xi$.}
 \label{fig:overlap_vs_u}
\end{figure}

We also computed the total overlaps $\mathcal{O}_\text{tot}=\frac{1}{d}\sum_{i,j=1}^d |\sand{\Psi_{GS,i}}{\Psi_{\mathrm{model},j}}|^2$  of the exact groundstates with the model state as a function of the interaction strength. The results are shown on Fig.~\ref{fig:overlap_vs_u}. One can notice that the overlaps are very high. Moreover, the overlap is an increasing function of $U$, as is the energy gap.

\begin{figure}[htb]
\includegraphics[width = 0.95 \columnwidth]{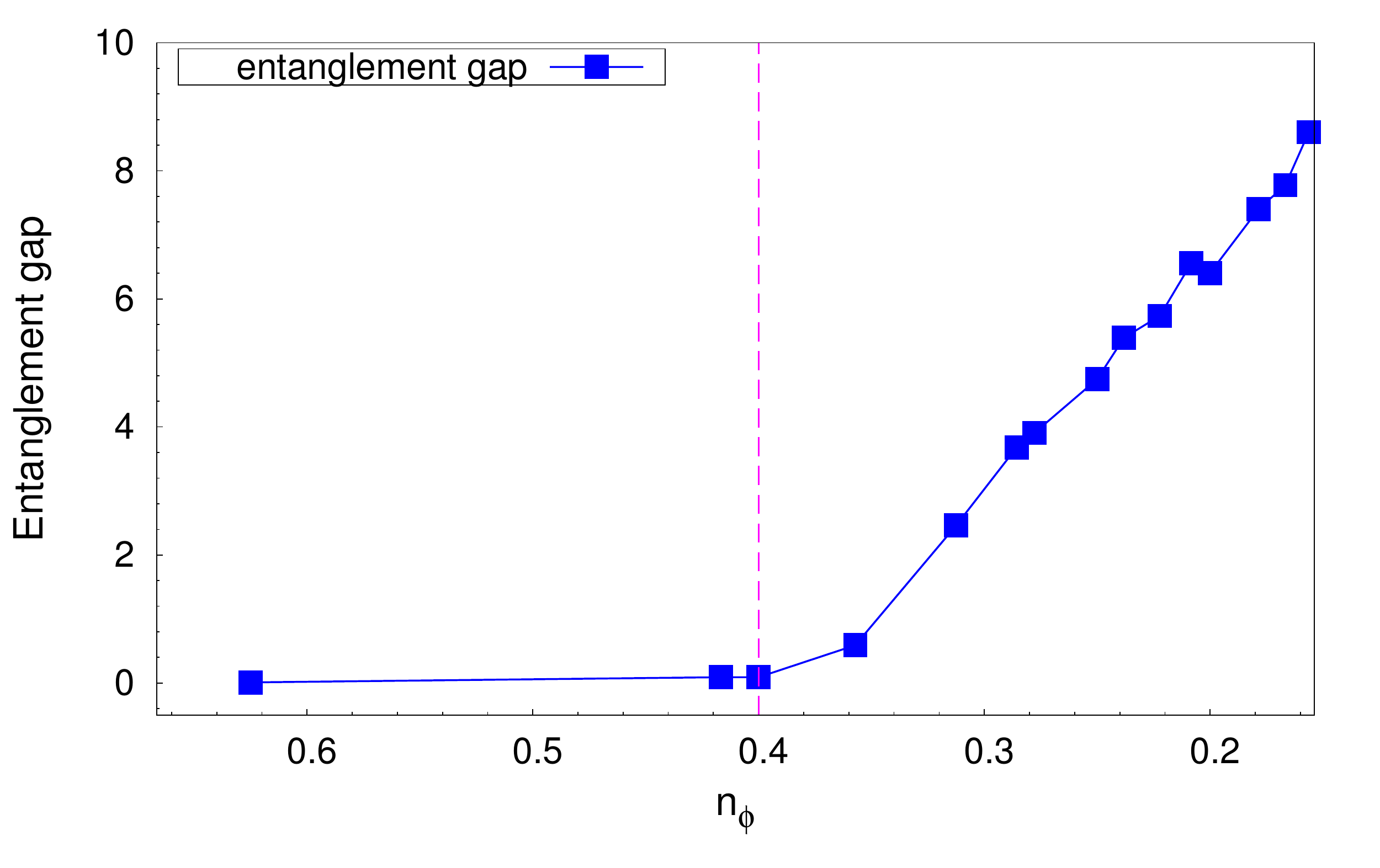}
\caption{Entanglement gap $\Delta_\xi$ as a function of flux density $n_\phi$ for the Laughlin state with $N = 5$ particles. The vertical dotted line indicates the  critical value $n_\phi^c\simeq 0.4$ up to which the groundstate exhibits the Chern number of the Laughlin state.\cite{Hafezi:2007p67}}
\label{fig:ES_gap_vs_size}
\end{figure}

It is now well established that the presence of an entanglement gap, in conjunction with the specific state counting in the PES, characterizes the topological order in the system. We are therefore interested to test how this measure compares to other signatures of topological order, such as the presence of non-zero Chern number for the groundstate manifold. A prior study of the Laughlin state on lattices had shown that the combined twofold ground state manifold has a Chern number of $1$, or $1/2$ per state, up to a critical flux density of $n_\phi^c\simeq 0.4$.\cite{Hafezi:2007p67} We now study how the entanglement gap varies as the flux density changes, by calculating entanglement spectra for systems of constant $N$ on lattices of different geometries. The results, shown in Fig.~\ref{fig:ES_gap_vs_size}, show a full agreement with Ref.~\onlinecite{Hafezi:2007p67}: For large $U$, we find that the entanglement gap also closes at $n_\phi^c\simeq 0.4$. However, while the Chern number jumps instantaneously between integers, the entanglement gap can capture how the topological protection of the Laughlin state is gradually weakened and finally collapses.

\subsection{Moore-Read State}

We next considered the Moore-Read (MR) state at $\nu = 1$. In the continuum, as explained in \ref{sec:MROnTorus}, the three MR states can be obtained from the Laughlin state by symmetrization.  On the lattice, the same scheme applies except that momentums are now defined modulo $K_y^\text{max}$, and we use the Laughlin states on the lattice as defined in Eqns.~(\ref{eq:Laughlin})-(\ref{eq:CM}) as the starting point.

\begin{center}
\begin{figure}[htb]
\includegraphics[width = 8 cm]{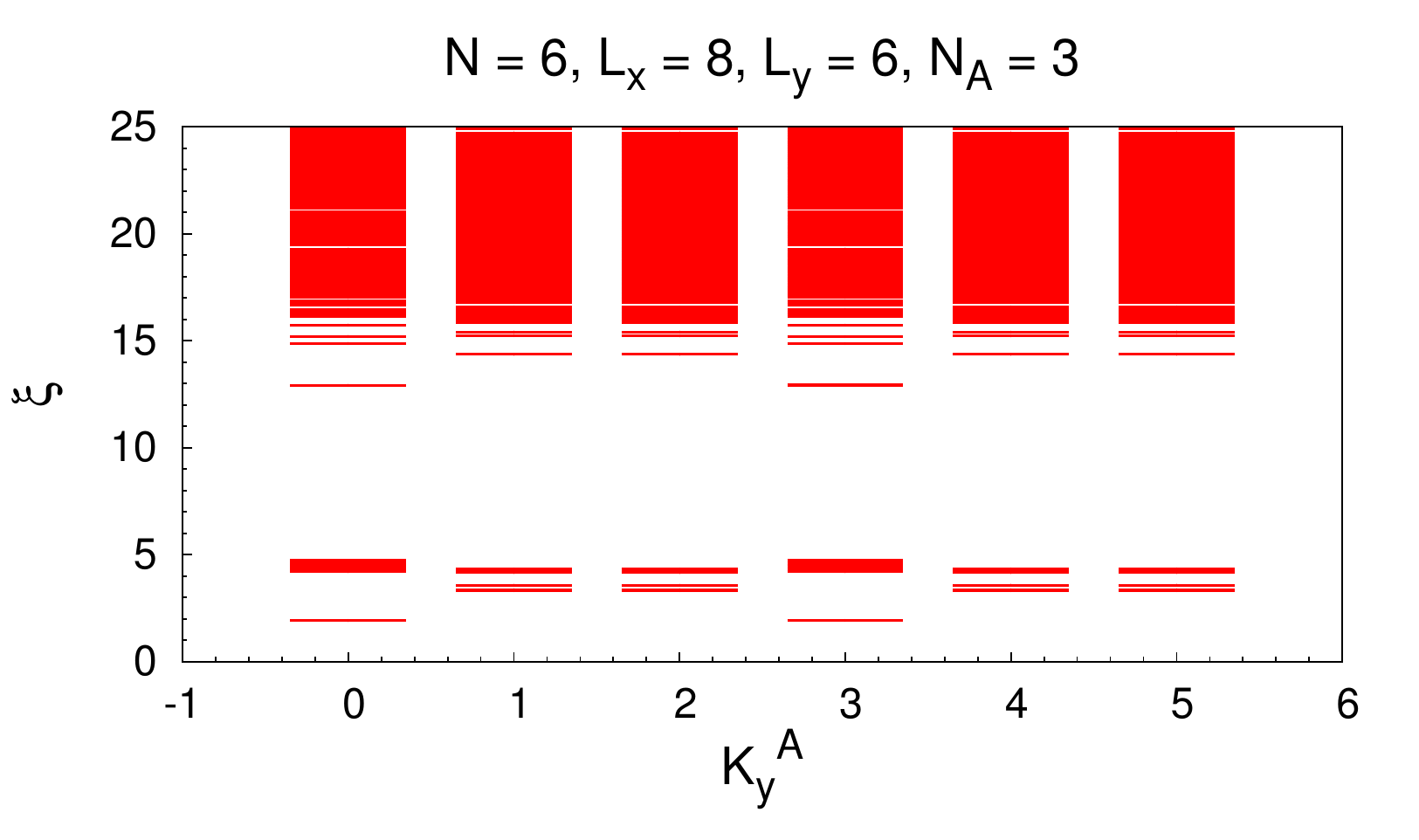}
\caption{Particle entanglement spectra for $N = 6$ bosons on a lattice at filling factor $\nu = N/N_\phi=1$, for a particle partition with $N_A = 3$ and for different lattice geometries. The spectra are calculated for the three-fold degenerate groundstate manifold of the Hamiltonian (\ref{eq:Hamiltonian}) with $U=0$ and $V/t=1$. Left: The counting of the states below the gap is $(7,6,6,7,6,6)$. It matches the one of MR quasiholes states on the torus as $K_y^\text{max} =  N_\phi$.}
\label{fig:MR_N6}
\end{figure}
\end{center}

Our numerical work on the lattice is based on the exact diagonalization of the Hamiltonian (\ref{eq:Hamiltonian}), using three-body contact interaction which are analogous to the continuum case, \ie choosing $U = 0$ and $V/t = 1$  in equation (\ref{eq:Hamiltonian}). Given these parameters, we generally found that the groundstate is approximately three-fold degenerate and the sectors in which the three ground states appear are given by the expected momenta, subject to the folding rule (\ref{eq:mappingN}). For geometries where the latter is satisfied, we compute the particle entanglement spectrum of the groundstates total density matrix. As in the 
Laughlin case, the particle entanglement spectrum is gapped and the number of states below the gap is given by the one predicted from the folding rule and the torus counting [see Fig.~\ref{fig:MR_N6}(a)]. The results for the different systems we studied are gathered in table \ref{tab:MR_system}.

\begin{table*}[htb]
\begin{tabular}{cccccccccc}
\hline
$N$ & $L_x$ & $L_y$ & $k_y(\text{GS})$  & match &$\Delta$ &  $\delta$ & PES: $\{\mathcal{N_A}_L(k_y)\}$ & match & $\Delta_\xi$\\
\hline 
4 & 5& 4& 0,0,2&\checkmark &0.016 &0.005 & $(3,2,3,2)$&\checkmark & 7.4\\
4&9&4& 0,0,2 &\checkmark& 0.015&4.2e-06&$(3,2,3,2)$&\checkmark  &14.2\\
4&10&4&0,0,2 &\checkmark& 0.005 &3.1e-04&$(3,2,3,2)$&\checkmark  &15.3\\
4 & 4& 6&0,0,0& \tickNo& & & & & \\
4 & 6& 6&0,0,0 &\checkmark &0.011 &0.0012 & $(6,4)$&\checkmark &11.8\\
4&8&6&0,0,0 &\checkmark&0.0063&1.7e-05&$(6,4)$&\checkmark&14.1 \\
4&10&6&0,0,0 &\checkmark&0.0036&1.0e-06&$(6,4)$&\checkmark&16.0\\
4&4&8&0,0,2&\checkmark&0.01&1.7e-05&$(3,2,3,2)$&\checkmark&13.0\\
4&6&8&0,0,2&\checkmark&0.006&1.7e-05&$(3,2,3,2)$&\checkmark&14.2\\
4&8&8&0,0,2&\checkmark&0.0042&1.3e-07&$(3,2,3,2)$&\checkmark&15.2\\
\hline
6 & 6& 4& 0,0,1&\checkmark & 0.044& 0.036 &$(19,19)$ & &0.39\\
6 & 8& 4& 0,0,1& \checkmark&0.017 & 0.0085& $(19,19)$& &0.38\\
6 & 6& 6& 0,0,3&\checkmark & 0.015& 1.3e-04&$(7,6,6,7,6,6)$ &\checkmark& 8.2 \\
6 & 8& 6& 0,0,3&\checkmark &9.4e-3 &3e-5 &$(7,6,6,7,6,6)$ &\checkmark & 8.2\\
\hline
\hline
\end{tabular}
\caption{Properties of the particle entanglement spectra of the groundstate manifold of the Hamiltonian for $U = 0$ and $V/t = 1$ (\ref{eq:Hamiltonian}) for different model systems of $N$ bosons with $N_\phi=N$ flux quanta. We indicate the momenta of the three degenerate groundstates $k_y(\text{GS})$ as well as the energy gap $\Delta$ and the groundstate energy splitting $\delta$. The counting of the particle entanglement spectrum is shown for the partition with $N_A = \lfloor N/2 \rfloor$ and for different lattice geometries. For both these properties, we indicate the agreement with the predictions for the Moore-Read state and the entanglement gap $\Delta_\xi$.}
\label{tab:MR_system}
\end{table*}

\begin{figure}[htb]
\includegraphics[width = 0.95 \columnwidth]{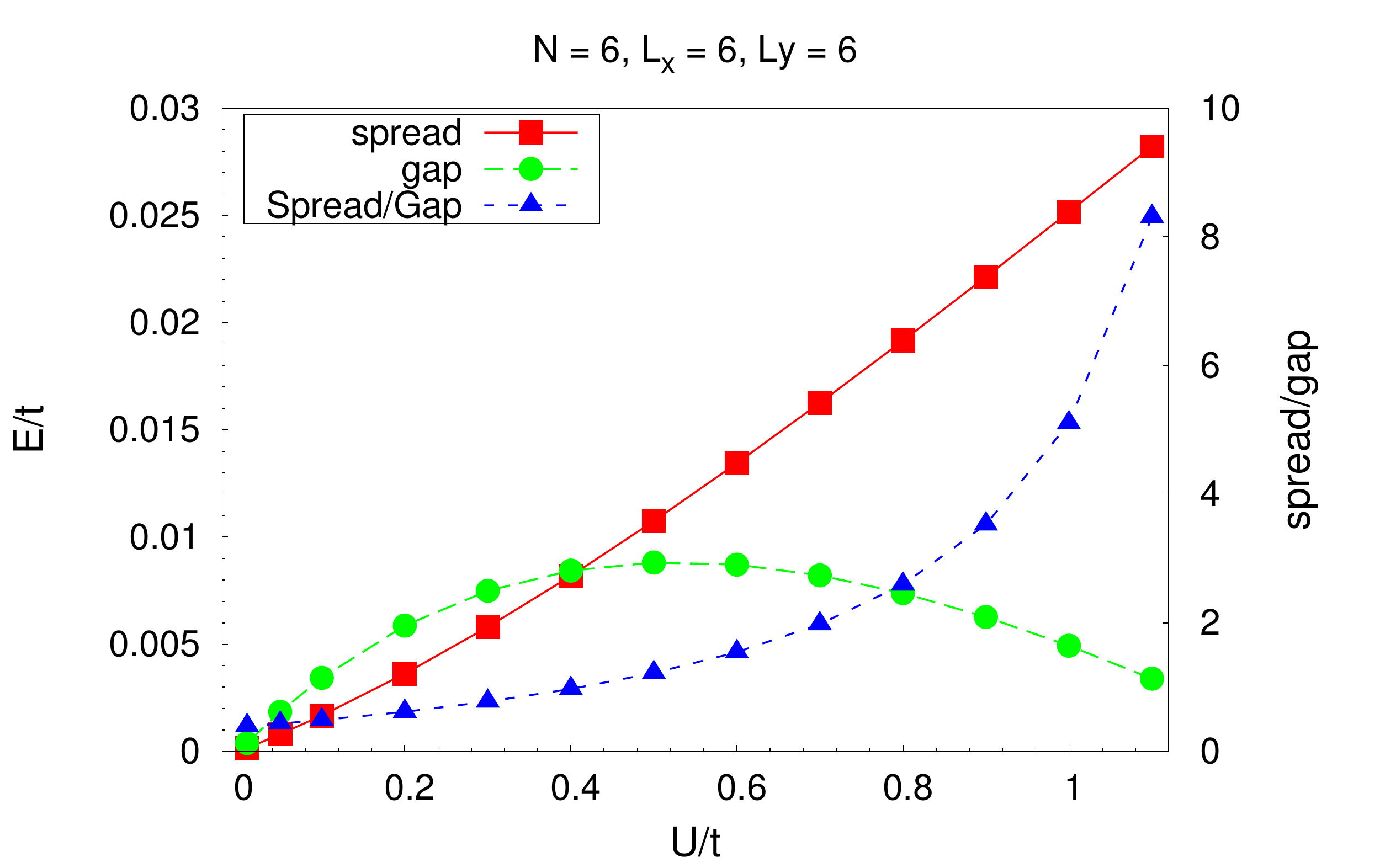}
\caption{Energy gap $\Delta=E_3-E_2$ and groundstate energy splitting $\delta=E_2-E_0$ as a function of $U$ for $\nu=1$, $N = 6$, $L_x = 6$ and $L_y = 6$, as well as their dimensionless ratio. }
 \label{fig:Egap_MR_nb_2_vs_u}
\end{figure}

These results show that it is theoretically possible to obtain the Moore-Read phase on the lattice using three-body contact interactions. Even though three-body interactions can be realized for cold atoms using more elaborate experimental settings,\cite{Syassen:2008p3058,Daley:2009p3057} the most relevant interaction for bosons on a lattice is the two-body hardcore interaction. Thus, we wonder if the Moore-Read state can also be stabilized with this type of interaction. In the continuum limit and in the lowest Landau level approximation, they are several numerical eveidences that such a phase can be stabilize.\cite{Cooper-PhysRevLett.87.120405,Regnault-PhysRevLett.91.030402,Chang-PhysRevA.72.013611,Regnault-PhysRevB.76.235324} Given the presence of LL mixing mixing in our model, we also establish to which extent this mixing affects the stability of the Moore-Read state. To answer these questions, we diagonalize the Hamiltonian (\ref{eq:Hamiltonian}) with $V = 0$ at $\nu = 1$. For small interaction strength $U$, the energy spectrum exhibits the correct groundstate degeneracy and -sectors. In addition, the entanglement spectra exhibit the correct state counting, albeit with a smaller entanglement gap than in the case of three-body interactions. 
A closer look at the spectrum reveals, as displayed in Fig.~\ref{fig:Egap_MR_nb_2_vs_u}, that the energy gap and the groundstate energy splitting are of the same order: While a fully realized MR phase should have a very small spread to gap ratio, this is clearly not the case here. 

We find that the gap closes for $U_c\simeq 1.25t$, while the spread between the ground states grows monotonically with $U$. Hence, the spread to gap ratio is a rapidly increasing function of $U$ indicating how sensitive the topological degeneracy is to the strength of two-body interactions. At the flux density $n_\phi=1/6$ shown in Fig.~\ref{fig:Egap_MR_nb_2_vs_u}, the lowest Landau level is still very narrow,\cite{Hofstadter:1976p69,Hormozi:2012p3156} so we interpret this strong suppression of the gap as resulting from Landau level mixing.  Furthermore, the single particle gap between the LLL and second LL is about $\Delta_{sp}\simeq 1.68 t$ at $n_\phi=1/6$, which is of the same order as the energy scale of interactions $nU_c$ at the point $U_c$ where the gap closes. We can conclude, at least for small systems, that the Landau level mixing resulting from large two-body contact interactions tends to destroy the MR phase. 

\begin{figure}[htb]
\includegraphics[width = 0.95 \columnwidth]{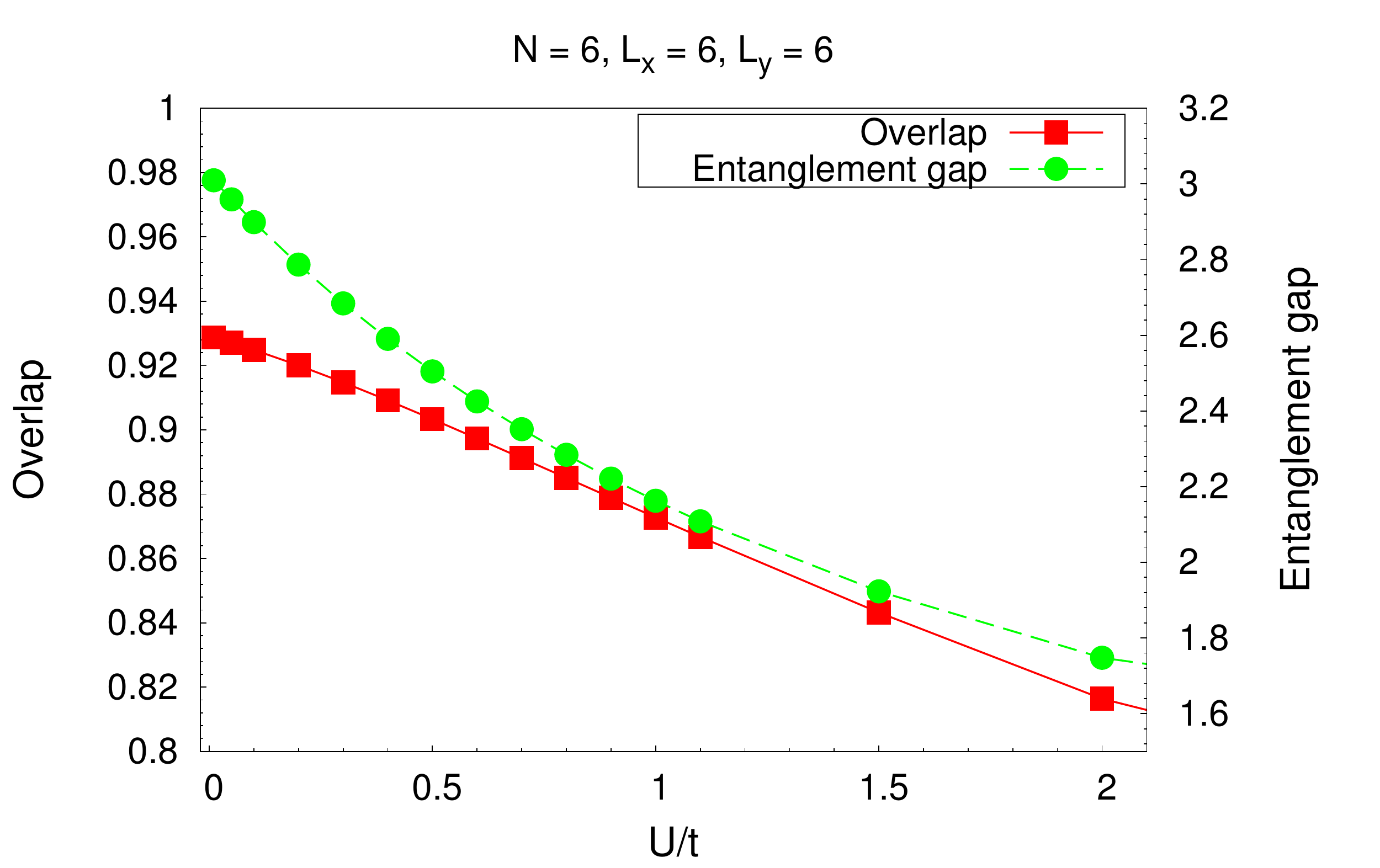}
\caption{Overlap $\mathcal{O}$ and entanglement gap $\Delta_\xi$ in the $N_A = 3$ sectors as a function of $U$ for $\nu=1$, $N = 6$, $L_x = 6$ and $L_y = 6$. Note the different offsets on the axes for the overlap $\mathcal{O}$ and entanglement gap $\Delta_\xi$.}
 \label{fig:overlap_MR_vs_u}
\end{figure}

Finally, we have computed the entanglement gap of the total matrix density, taking the two lowest energy state in the $K_y = 0$ sector and the lowest in the $K_y = N/2$ sector even when they were not the three lowest energy eigenstates. For reference, we also evaluate the overlap of these states with the model MR states, as discussed above. The results, shown in Fig.~\ref{fig:overlap_MR_vs_u}, are consistent with the previous conclusions: The phase obtained at small $U$ is most likely the MR phase, with large overlaps at small $U$, but the phase is destroyed by increasing the interaction strength.

\section{Conclusion}
\label{sec:Conclusions}

In this paper we have analyzed the bosonic fractional quantum Hall states on lattices through the particle entanglement spectrum. These systems provide a well controlled environment away from the pure model states, which allows a better understanding of the properties of particle entanglement spectra (PES) in quantum Hall systems. We have focused on the filling factors $\nu=\frac{1}{2}$ and $\nu=1$ where the Laughlin state and the Moore-Read state should respectively emerge. In both cases, the PES was able to discriminate the nature of the state. This result is even more remarkable given that the size of the Hilbert space, set by the number of particles and lattice sites (rather than flux quanta), is exponentially larger than in the continuum limit. Interestingly, the PES was able to give insight about a competing Bose-Einstein condensate phase, which we have associated with low lying entanglement eigenstates whose number is invariant under the number of particles in the partition. We have also shown that the entanglement gap collapse in the PES predicts a critical density of flux $n_\phi^c$ below which the Laughlin's physics emerges; our value of $n_\phi^c$ is in agreement with a previous study based on Chern numbers.  We have used the PES to confirm the realization of a Moore-Read state at $\nu=1$ in the presence of on-site two-body contact interactions, only. Furthermore, we have given evidence of how Landau level mixing arising from these two-body contact interactions tends to destroy the bosonic MR state as its magnitude is increased.
 
\begin{acknowledgments}
We acknowledge M. Hermanns for useful discussions. A.S. thanks Princeton University for generous hosting. A.~S. was supported by Keck grant. N.~R. was supported by  NSF CAREER DMR-095242, ONR - N00014-11-1-0635, Packard Foundation and Keck grant. This material is based upon work supported in part by the National Science Foundation Grant No. 1066293 and the hospitality of the Aspen Center for Physics. G.M. gratefully acknowledges support from the Leverhulme Trust under grant ECF-2011-565 and from the Newton Trust of the University of Cambridge.
\end{acknowledgments}

\bibliography{LatticePes}

\end{document}